\documentclass[twocolumn,preprintnumbers,amsmath,amssymb,superscriptaddress]{revtex4}
\usepackage{graphicx}
\usepackage{dcolumn}
\usepackage{bm}
\usepackage{soul}
\usepackage{color}
\usepackage{epstopdf}
\usepackage[version=3]{mhchem}
\usepackage{lipsum}
\usepackage[outercaption]{sidecap}
\usepackage{floatrow}
\usepackage{hyperref}

\begin{document}


\title{Tailoring Drug Mobility by Photothermal Heating of Graphene Plasmons}

\author{Anh D. Phan}
\email{anh.phanduc@phenikaa-uni.edu.vn}
\affiliation{Faculty of Materials Science and Engineering, Phenikaa University, Hanoi 12116, Vietnam}
\affiliation{Phenikaa Institute for Advanced Study, Phenikaa University, Hanoi 12116, Vietnam}
\author{Nguyen K. Ngan}
\email{ngan.nguyenkim@phenikaa-uni.edu.vn}
\affiliation{Faculty of Materials Science and Engineering, Phenikaa University, Hanoi 12116, Vietnam}
\author{Do T. Nga}
\affiliation{Institute of Physics, Vietnam Academy of Science and Technology, 10 Dao Tan, Ba Dinh, Hanoi 12116, Vietnam}
\author{Nam B. Le}
\affiliation{School of Engineering Physics, Hanoi University of Science and Technology, 1 Dai Co Viet, Hanoi 10000, Vietnam}
\author{Chu Viet Ha}
\affiliation{Faculty of Physics, Thai Nguyen University of Education, Thai Nguyen 24000, Vietnam}
\date{\today}

\begin{abstract}
We propose a theoretical approach to quantitatively determine the photothermally driven enhancement of molecular mobility of graphene-indomethacin mixtures under infrared laser irradiation. Graphene plasmons absorb incident electromagnetic energy and dissipate them into heat. The absorbed energy depends on optical properties of graphene plasmons, which are sensitive to structural parameters, and concentration of plasmonic nanostructures. By using theoretical model, we calculate temperature gradients of the bulk drug with different concentrations of graphene plasmons. From these, we determine the temperature dependence of structural molecular relaxation and diffusion of indomethacin and find how the heating process significantly enhances the drug mobility.
\end{abstract}

\keywords{Suggested keywords}
\maketitle
Photothermal effects have a wide range of applications such as solar steam generation \cite{17}, thermo-optical data writing \cite{18}, cancer treatment \cite{21,22}, and destroying bacteria \cite{23}. The incident electromagnetic field excites collective oscillations of free electrons, which are surface plasmons in metal-like materials. Plasmonic excitations allow confining light near the metal-dielectric interface and enhancing local fields. Thus, plasmonic materials can absorb light and turn it into heat. Investigating mechanisms underlying light-to-heat conversion not only provides insight into fundamental science, but also proposes novel practical applications. 

Very recently, photothermal plasmonic nanoparticles have been experimentally exploited to induce drug amorphization inside a table \cite{8} upon laser irradiation. Compared with other hyperthermia methods (using thermal chambers, ultrasound, microwave, and magnetic nanoparticles associated with an alternating magnetic field) \cite{29}, the photothermal approach has a simple operation and an advantage to locally and selectively increase temperature instead of spreading the heat to larger zones. Compared to global heating, a localized heating effect controls on-demand release of drugs in abnormal cell regions at a desired time and location to improve pharmaceutical efficacy. In addition, global and uniform heating can damage to surrounding normal tissues and induce direct cell necrosis. Thus, it is a highly efficient technique and side effects are minimized. The laser heating increases temperature and molecular mobility of the drug and polymer in the tablet. Molecular mobility is characterized by molecule's structural relaxation and diffusion, which can be measured using the broadband dielectric spectroscopy. The photothermal heating of drugs may be an answer to overcome poor aqueous drug solubility in oral drug delivery \cite{24,25}. Recrystallization of pharmaceuticals during manufacturing or storage, which restricts the use of amorphous drugs, can be resolved by plasmonic heating. All challenging problems are of interest for the scientific and industrial community but have not been theoretically investigated.

Among many types of photothermal agents, graphene has received much attention since it has various intriguing behaviors \cite{2,3,4,20}. Graphene exhibits strong light absorption with relatively low losses \cite{20}. The number of free electrons in graphene is small and this leads to a significant reduction of the heat dissipation compared to noble metals \cite{20}. Large inelastic losses reduce lifetimes of plasmon and optical confinement. It is easy to tune plasmonic properties of graphene by using doping, defects, external fields, and charge injection \cite{4}. 

Although graphene nanostructures are composed of carbon atoms which is more friendly with human body than conventional plasmonic materials, their toxicity still exists and strongly depends on shape, size, purity, concentration, and dose \cite{30,31,32}. For example, wistar rats have no sign of toxicity when injected doses of 1 mg/kg and 50 mg/kg of 100-nm graphene nanodisks, however, a 100 mg/kg dose killed two of eight rats after 2 weeks of injection \cite{33}. Meanwhile, graphene nanostructures having size in the range of 500-1000 nm can cause cytotoxicity when studying the viability of RAW 264.7 macrophages injected by 20-100 $\mu g/ml$ \cite{34}. Oxidation of graphene was found to be a main reason for toxicity in the lungs of mice \cite{35}. According to guidance for safety studies of Food and Drug Administration (FDA) \cite{36}, a maximum safe starting dose is 100 $\mu g$ per subject for clinical trials and the maximum dose 100 times larger than the clinical dose can be used for nonclinical studies. Previous \emph{in vivo} experiments \cite{33, 35, 37, 38} clearly confirmed the safety of the recommended clinical dose. Using higher doses can cause either several serious problems in some cases or minor side effects \cite{37, 38}. Overall, graphene nanostructures having the 100-nm size possibly have less toxicity than their larger counterparts and health risks can be potentially found when investigated at the dose exceeding 100 $\mu g$ per subject. The toxicity is governed by the complex interplay of physiochemical properties but it remains not well understood and further experimental investigations are necessary. This is why graphene-based nanomaterials have been received much attention to improve the biocompatibility and propose various new biological and pharmaceutical applications.

\begin{figure}[htp]
\includegraphics[width=8cm]{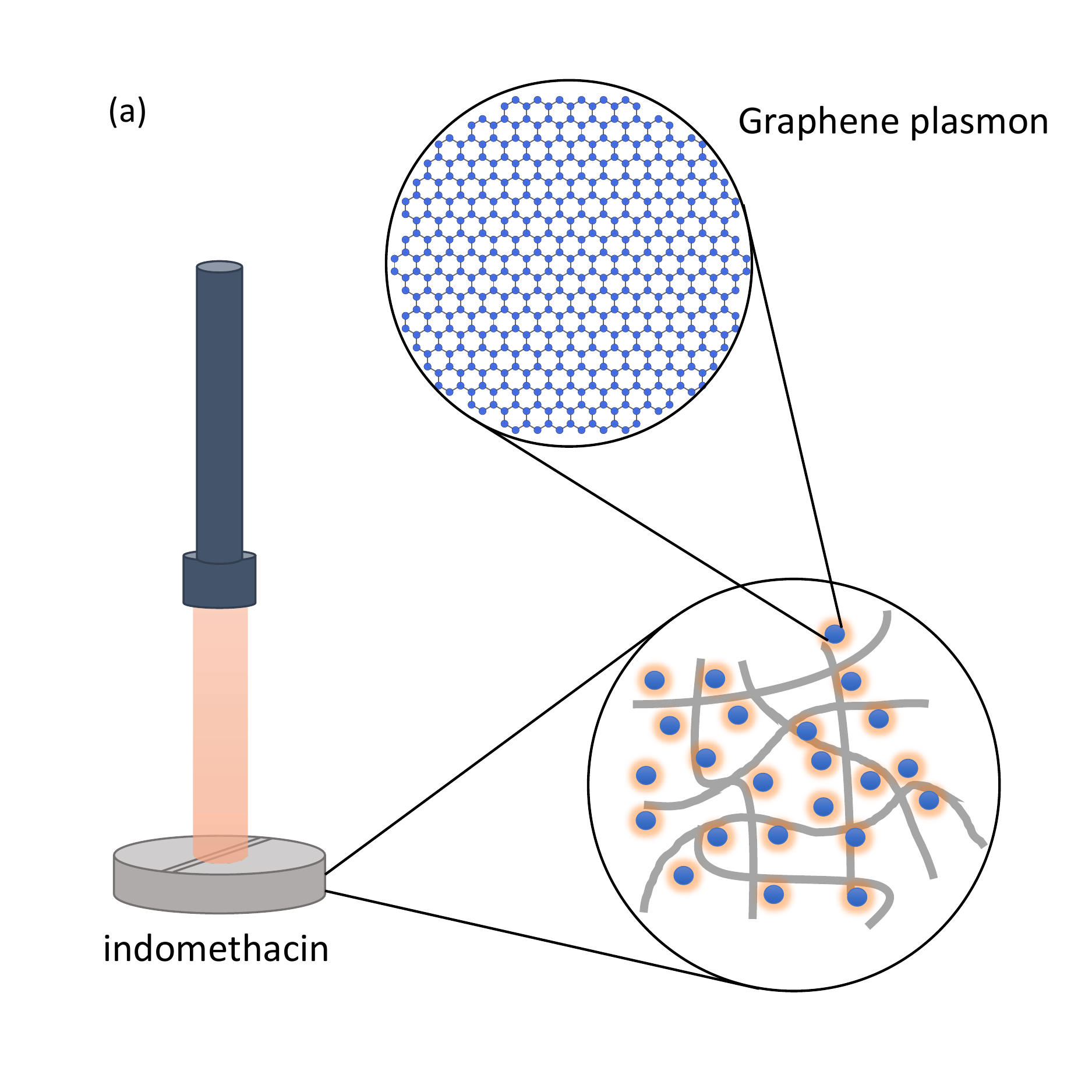}
\includegraphics[width=8cm]{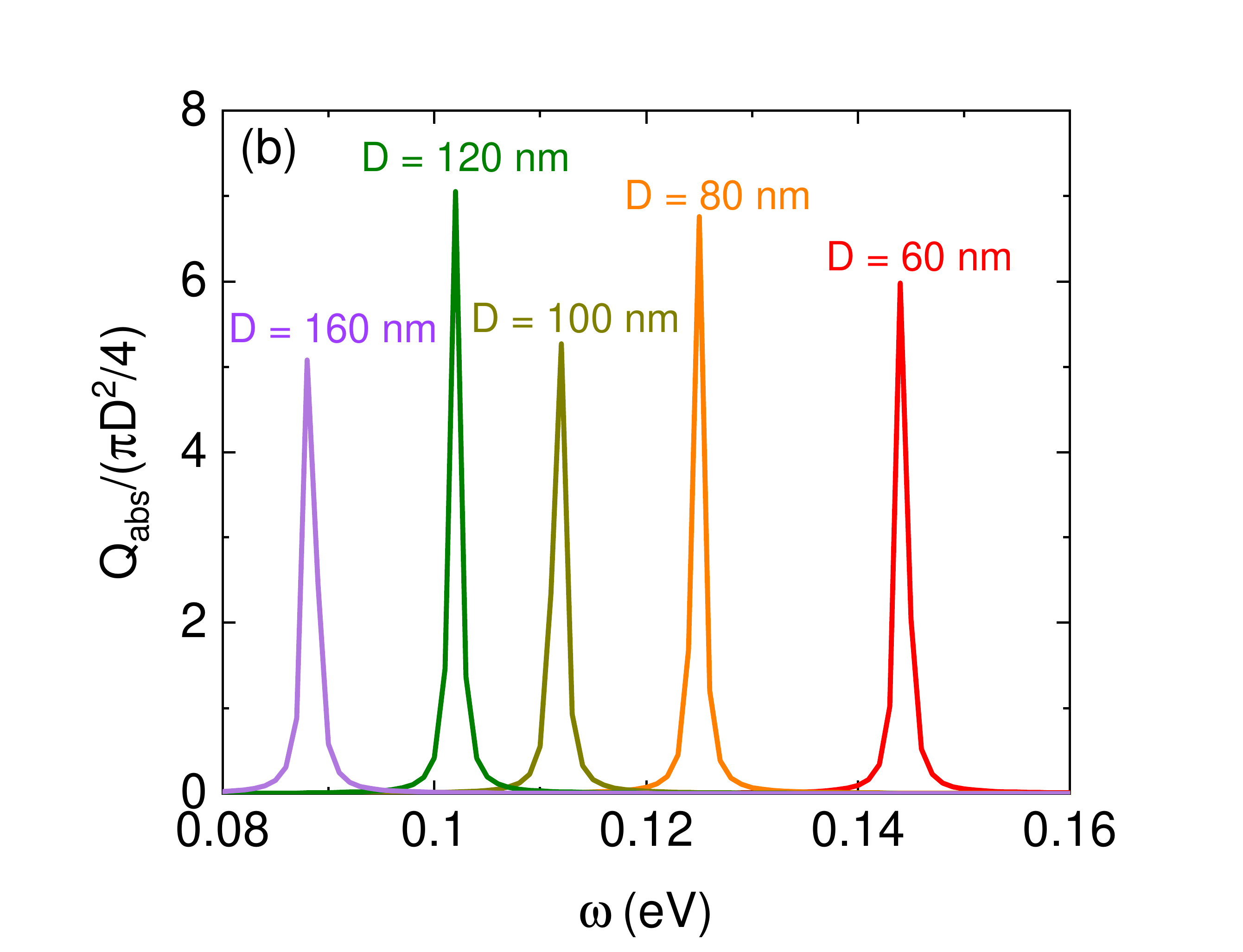}
\includegraphics[width=8cm]{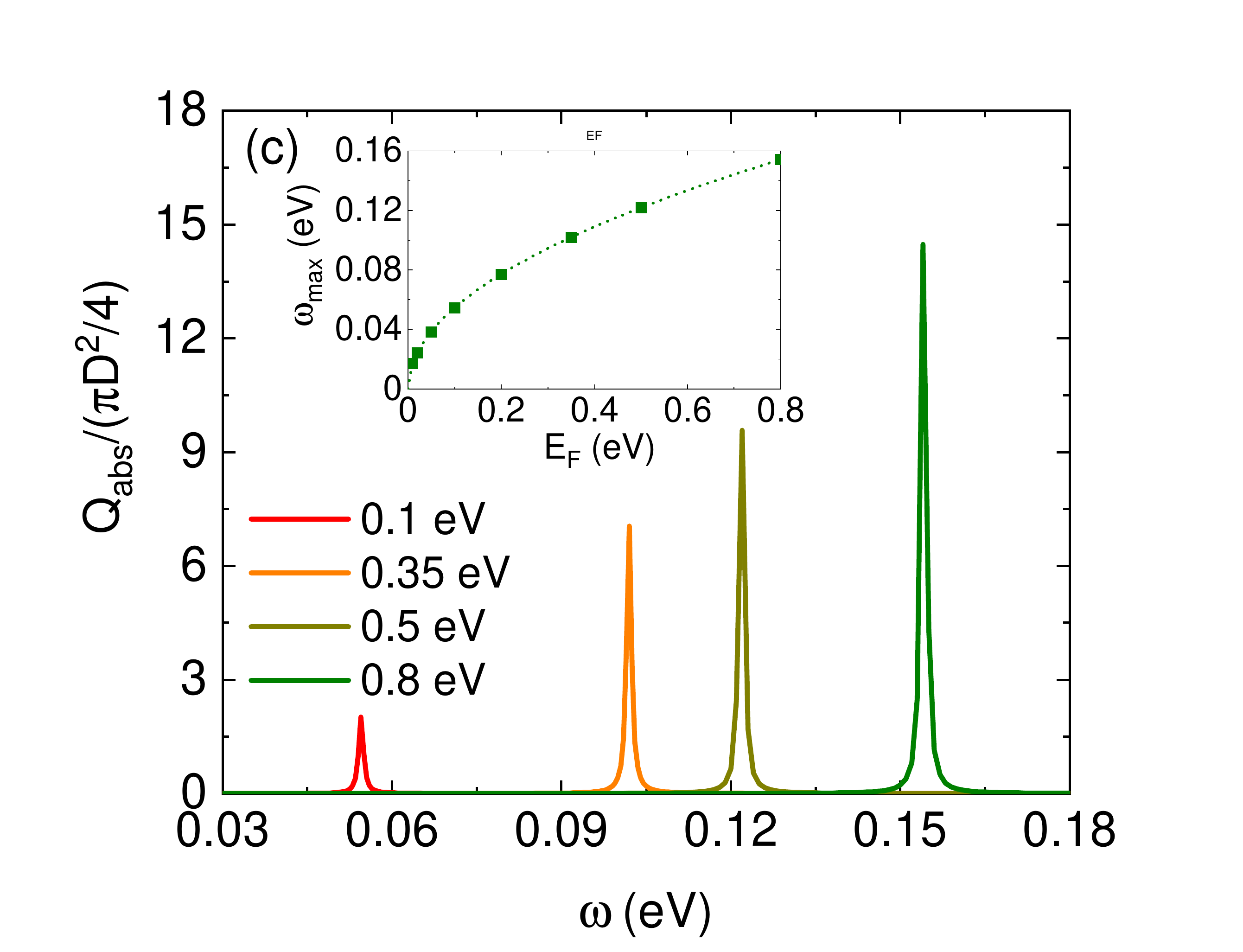}
\caption{\label{fig:1}(Color online) (a) Schematic illustration of graphene-indomethacin mixture exposed to laser radiation. The frequency dependence of the absorption cross section of an isolated graphene plasmon normalized by its area at different (b) diameters $D$ and (c) Fermi energies for $D = 120$ nm. The inset of (c) presents the resonant peak as a function of $E_F$. The data points and dotted curve correspond to full calculations (Eq. (\ref{eq:1})) and fit function ($\omega_{max}\approx0.17\sqrt{E_F}$), respectively.}
\end{figure}

In this letters, we theoretically investigate effects of photothermal heating on molecular mobility in a graphene-drug composite under a infrared laser illumination (Fig. \ref{fig:1}a). Monolayer graphene plasmons are randomly dispersed in bulk indomethacin and considered as heat sources. Absorption and heat generation of this mixture depends on optical properties and density of graphene nanostructures. After determining the temperature dependence of structural relaxation and diffusion by the elastically collective nonlinear Langevin equation (ECNLE) theory \cite{11,12,13}, we calculate time-dependent temperature of the drug mobility upon laser irradiation.

When the wavelength of the incident electromagnetic field is much larger than the size of graphene plasmon, the quasi-static approximation can be exploited to analytically determine the absorption cross section of a single graphene plasmon \cite{1,15,4,5} by
\begin{eqnarray}
Q_{abs}=\frac{4\pi\omega}{c}Im\left[\frac{\sqrt{\varepsilon_m}D^3\zeta^2}{-i\omega D\cfrac{\varepsilon_m}{\sigma(\omega)}-\cfrac{1}{\eta}}\right],
\label{eq:1}
\end{eqnarray}
where $\varepsilon_m$ is the dielectric constant of medium, $\omega$ is the frequency of the incident field, $D$ is the diameter of graphene nanodisks, $\sigma(\omega)$ is the optical conductivity, and $\zeta = 0.8508$ and $\eta=-0.0728$ are morphological parameters \cite{1}. In mid-infrared regime, the monolayer graphene conductivity is \cite{4}
\begin{eqnarray}
\sigma(\omega)=\frac{e^2 i |E_F|}{\pi\hbar^2\left(\omega + i\tau^{-1} \right)},
\label{eq:2}
\end{eqnarray}
where $E_F$ is a chemical potential, $e$ is the charge of electron, $\hbar$ is the reduced Planck constant, and $\tau$ is the carrier relaxation time. We choose $E_F=0.35$ eV and $\hbar\tau^{-1}=0.001$ eV to be consistent with Ref. \cite{1}. Typically, the full expression of $\sigma(\omega)$ has two terms corresponding to interband and intraband transitions. However, the interband conductivity is extremely small in the mid-infrared regime. Thus, we ignore contributions of interband transitions in Eq. (\ref{eq:2}).

Figure \ref{fig:1}b shows influence of finite-size effects of graphene plasmon on the normalized absorption cross section calculated by Eqs. (\ref{eq:1}) and (\ref{eq:2}). Since there is no dielectric data for indomethacin the frequency range of 0.08-0.16 eV, we extrapolate dielectric data in Ref. \cite{10}, which is measured in the frequency range of $10^{-2}$-$10^7$ Hz. This extrapolation suggests that dielectric loss at higher frequencies is very small while the dielectric constant is $\varepsilon_m \approx 1.77$. Thus, the absorption of the drug is extremely lowered and we assume to ignore it in calculating $Q_{abs}$. The maximum values of the normalized absorption spectrum are approximately 5-7 although the size of graphene nanodisk increases from 60 nm to 160 nm. Graphene plasmons in this size range are safe for living subjects \cite{33,38}. One can roughly estimate the maximum absorption of graphene plasmons via $4Q_{abs}^{peak}/(\pi D^2)\approx6$. The resonant peak ($\omega_{max}$) is red-shifted when increasing $D$. In Ref. \cite{4}, Abajo provided a simple expression to connect the plasmon resonance position of graphene plasmons, its size and Fermi energy, which is $\omega_{max}\sim \sqrt{E_F/D}$. The inset of Fig. \ref{fig:1}c numerically confirms this relation for graphene plasmons of $D=120$ nm. Our numerical results based on Eq. (\ref{eq:1}) are perfectly fitted with a function of $\omega_{max}\approx0.17\sqrt{E_F}$. The Fermi energy is controlled by applying external electric field \cite{26} or doping \cite{1,4} which gives $E_F=\hbar v_F\sqrt{n}$, where $v_F\approx 10^6$ m/s is Fermi velocity in graphene and $n$ is the carrier density.

The dependence of the absorption spectrum on Fermi energy can be seen in the mainframe of Fig. \ref{fig:1}c. The spectrum is lowered when $E_F$ is decreased. The magnitude of absorption at resonances is analytically described by $4Q_{abs}^{peak}/\pi D^2\approx20.2E_F$ and this expression nearly remains unchanged with size since changing size does not change the ratio of $4Q_{abs}^{peak}/(\pi D^2)$ at a fixed Fermi energy.

Now, graphene plasmons having the density, $N$ (the number of nanostructures per volume), are randomly mixed with indomethacin drugs. In this work, we choose $N=5\times10^{13}-3\times10^{14}$ nanodisks/$\ce{m^3}$. The averaged separation distance between two adjacent graphene plasmons is $1/N^{1/3}\approx 14.9-27$ $\mu m \gg D$. It means that nanostructures in the drug are fully isolated and no plasmonic coupling occurs. Since Eq. (\ref{eq:1}) allows us to determine the in-plane optical absorption, we can effectively include effects of random orientation of graphene nanodisks on the graphene absorption by replacing $Q_{abs}$ with $2Q_{abs}/3$. From this, according to, an effective absorption coefficient of our binary mixture constructed using the Beer-Lambert law is $\alpha(\omega)=2NQ_{abs}/3$. %

Photothermal effects of the plasmonic agents under irradiation of infrared laser light are exploited to control drug mobility. We consider the photothermal heating of bulk indomethacin of graphene plasmon with $D=120$ nm when exposed by an infrared laser beam operating at $\sim 0.1$ eV (or 806.6 \ce{cm^{-1}}) and the intensity $I_0=1$ $W/cm^2$. Electromagnetic radiation passing through these systems is absorbed by the plasmonic materials and converts from optical energy to thermal energy to increase the temperature of the surrounding medium. The light-to-heat conversion process is described using heat transfer equations without taking into account radiative and convective heat losses.

To calculate the spatial temperature profile in the bulk indomethacin of graphene plasmons under laser illumination, we use an analytic expression of thermal response of a semi-infinite substrate \cite{6,7} derived from a heat energy balance equation, which gives
\begin{widetext}
\begin{eqnarray}
\Delta T(x,y,z,t)=\frac{I_0(1-\mathcal{R})\alpha}{2\rho_d c_d}\int_0^t \exp\left(-\frac{\beta^2(x^2+y^2)}{1+4\beta^2\kappa t'} \right)\frac{e^{\alpha^2\kappa t'}}{1+4\beta^2\kappa t'}\left[e^{-\alpha z}\ce{erfc}\left(\frac{2\alpha\kappa t'-z}{2\sqrt{\kappa t'}} \right)+ e^{\alpha z}\ce{erfc}\left(\frac{2\alpha\kappa t'+z}{2\sqrt{\kappa t'}} \right)\right]dt',
\label{eq:3}
\end{eqnarray}
\end{widetext}
where $z$ is the depth direction parallel with the incident field, $x$ and $y$ are coordinates in the horizontal plane, $\mathcal{R}$ is the reflectivity of indomethacin, and $\beta$ is the inverse of the laser spot radius. Suppose that the spot size is infinite, we have $\beta = 0$. The thermal diffusivity of the drug, $\kappa = K_d/\rho_d c_d$, is calculated by the thermal conductivity, $K_d=0.18$ W/m/K \cite{9}, the mass density, $\rho_d=1213$ \ce{kg/m^3}, and the specific heat capacity, $c_d=1200$ \ce{J/kg/K} \cite{9}. Equation (\ref{eq:3}) has been used to quantitatively describe the photothermal heating of a solution of TiN nanoparticles and provide a good agreement with experiments \cite{6}. We assume that the validation is still held in our calculations.

\begin{figure}[htp]
\includegraphics[width=8cm]{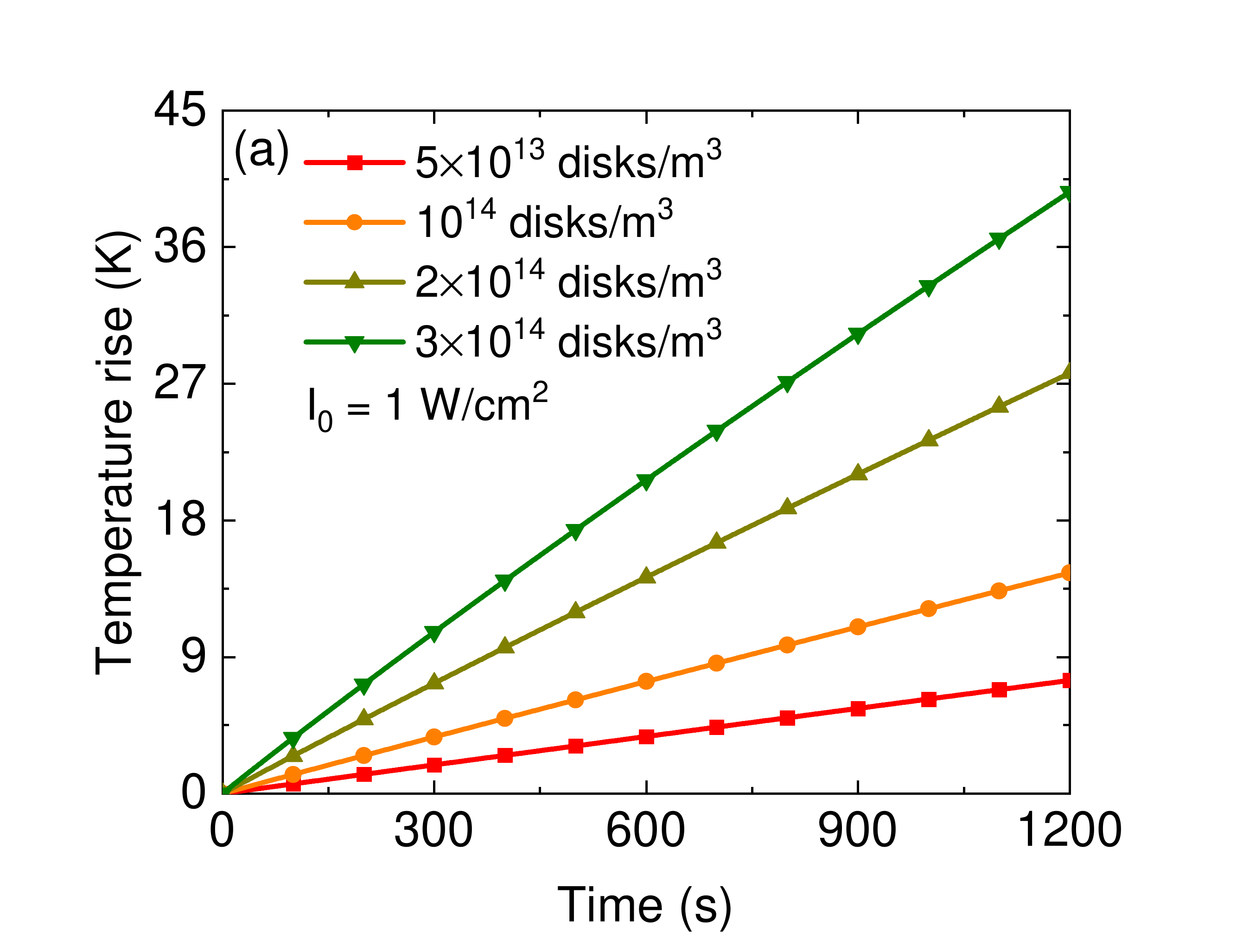}
\includegraphics[width=8cm]{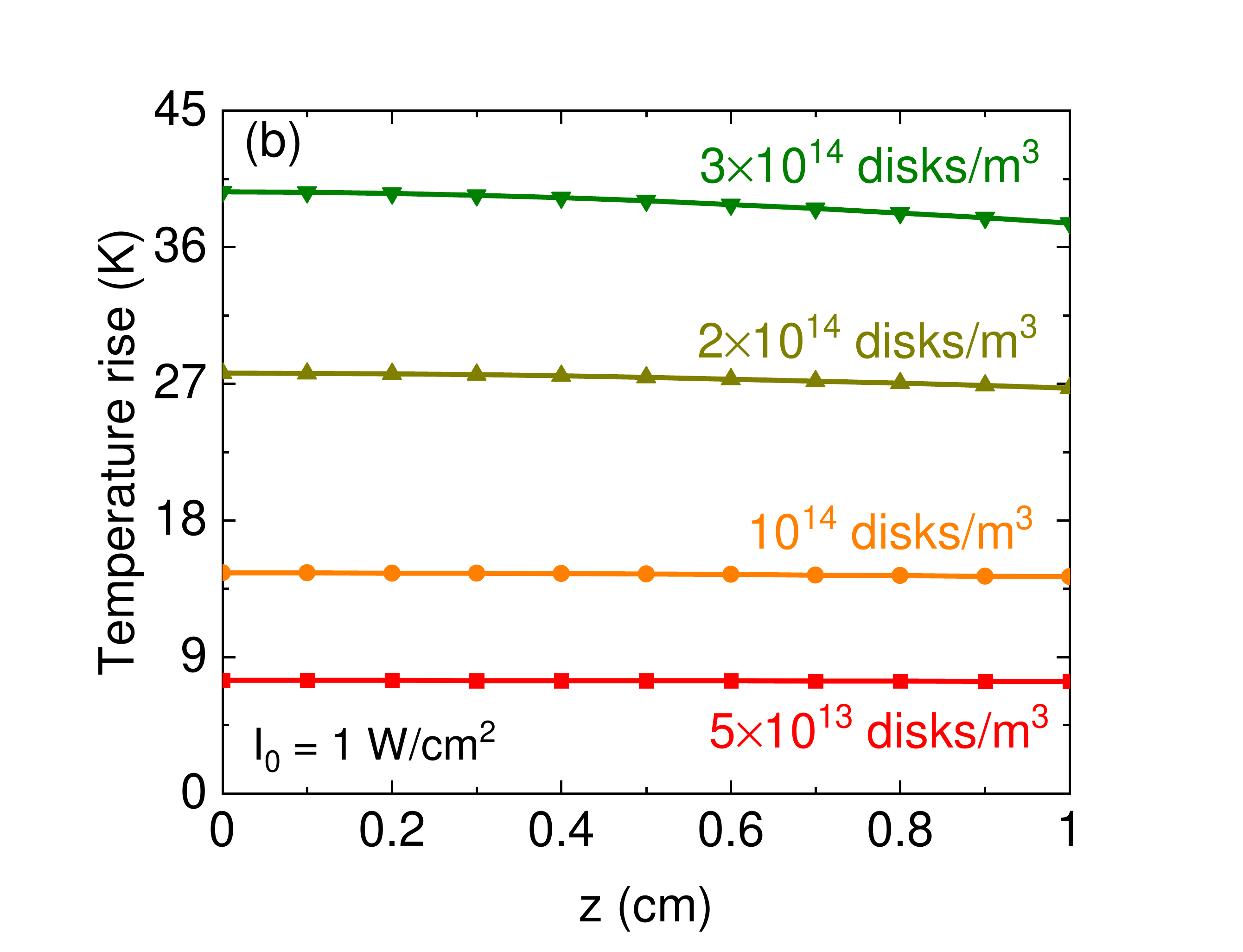}
\caption{\label{fig:2}(Color online) (a) Time dependence of the temperature rise at the drug surface ($z=0$) and (b) gradient of temperature rise at $t = 1200$ $s$ as a function of $z$ with different densities of graphene plasmons.}
\end{figure}

Based on Ref. \cite{27,28}, the transmission of indomethacin at 806.6 \ce{cm^{-1}} is about 97 $\%$. It suggests that $\mathcal{R}$ and the absorption of pure indomethacin are relatively small. Thus, we assume $\mathcal{R}\approx 0$ in Eq. (\ref{eq:3}) and the photothermal heating of graphene-drug composites are mainly caused by the absorbed energy in graphene plasmons. This is consistent with our assumption given to calculate the absorption cross section of graphene plasmons in Fig. \ref{fig:1}. The same performance ($\mathcal{R}\approx 0$) can be obtained when the frequency of laser and $\omega_{max}$ lie in the range of 1800-2500 \ce{cm^{-1}} \cite{27,28} or 0.22-0.31 eV. It means that the Fermi energy varies from 1.3 eV to 1.8 eV if the diameter of graphene plasmon is fixed at 120 nm. One can also reduce $D$ to shift the plasmonic resonance to this desired range of frequency while $E_F$ remains small. 

Figure \ref{fig:2} shows numerical results for the temporal and spatial temperature of the graphene-drug mixtures calculated using Eq. (\ref{eq:3}). Increasing the density of graphene plasmons not only leads to more absorption of optical energy in the mixture, but also shortens the penetration depth of electromagnetic fields. Thus, the absorbed energy is localized near the surface, and the light-to-heat conversion decays towards the drug center. This finding explains why the surface temperature monotonically grows with an increase of $N$ from $5\times10^{13}$ to $3\times10^{14}$ \ce{disks/m^3}, while the temperature in the drug interior slightly decreases in comparison with that in the surface. Changing temperature changes molecular mobility in the drug.

To investigate the temperature dependence of structural relaxation and diffusion of amorphous materials, we use the ECNLE theory. In this theory, all glass-forming liquids are modeled as a hard-sphere fluid of the volume fraction $\Phi$. A sphere of the fluid has the particle diameter, $d$. Molecular mobility of a tagged particle is affected by two main factors: (i) the nearest-neighbor interactions and (ii) cooperative motions beyond nearest neighbours. The former factor governs the local dynamics quantified by the dynamic free energy $F_{dyn}(r)=F_{ideal} (r)+F_{caging}(r)$ with $r$ being the displacement. $F_{ideal}(r)$ and $F_{caging}(r)$ correspond to the ideal fluid and localized state, respectively. The caging force acting on the tagged particle tightly depends on the density and structure of systems. 

\begin{figure}[htp]
\includegraphics[width=8cm]{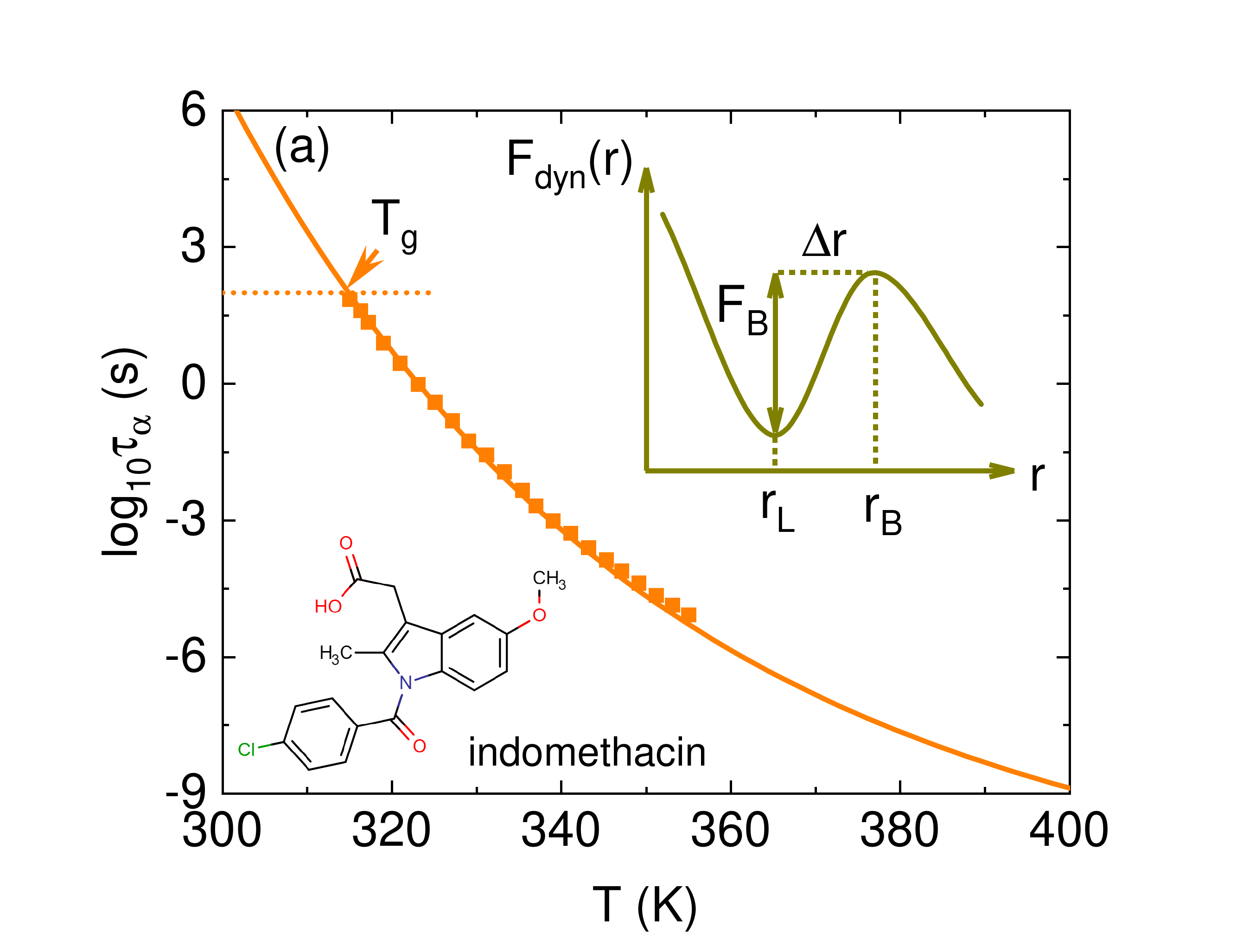}
\includegraphics[width=8cm]{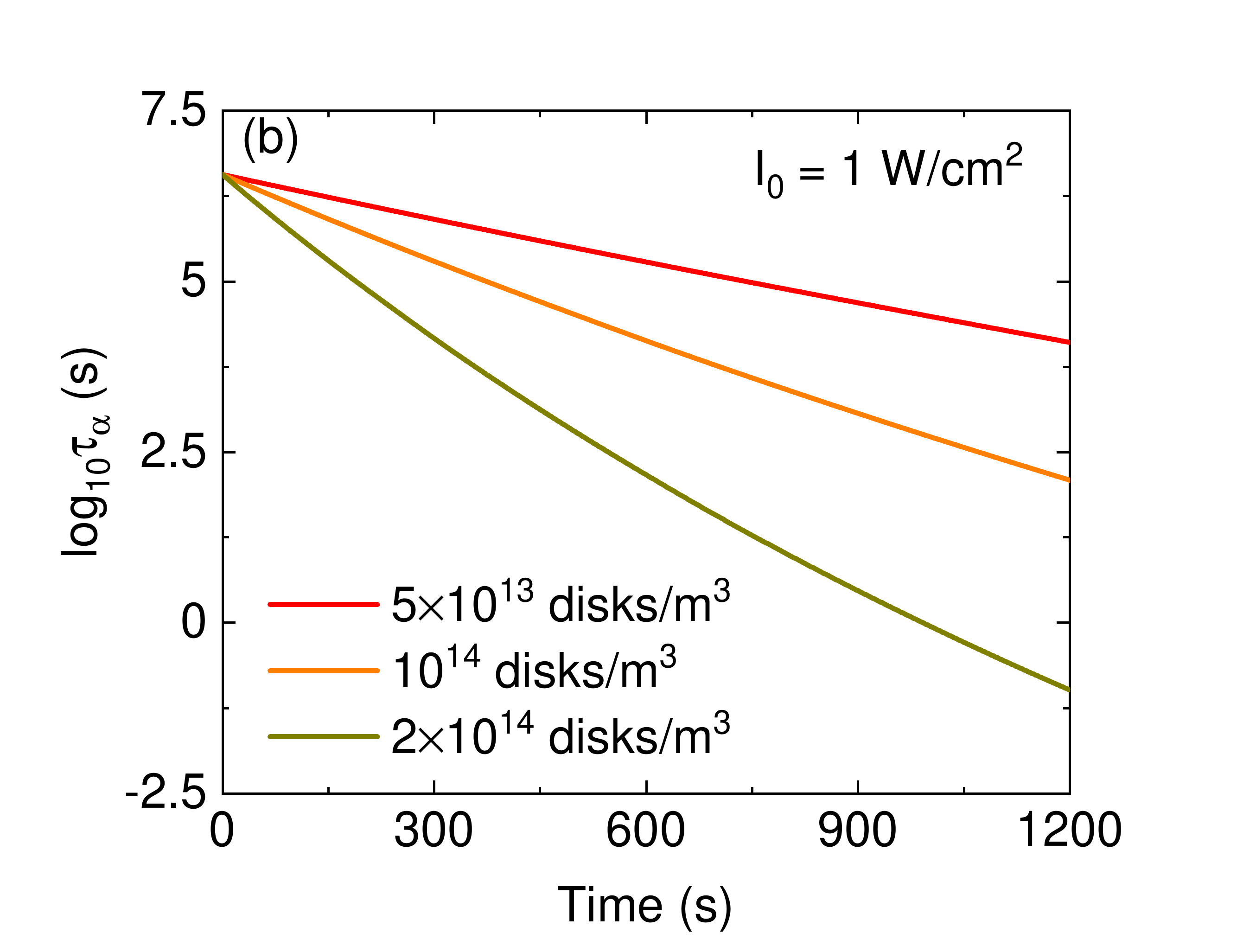}
\caption{\label{fig:3}(Color online) (a) Structural relaxation time of graphene-indomethecin composites as a function of (a) temperature without laser illumination and (b) time at $T = 300K$ under laser illumination with different densities of graphene nanodisks. Data points and solid curves correspond to experimental data \cite{14} and theoretical calculations, respectively. The inset shows illustration of $F_{dyn}(r)$ and characteristic parameters of the local dynamics.}
\end{figure}

The inset of Fig. \ref{fig:3}a illustrates the function form of $F_{dyn}(r)$ when the density is sufficiently large and the dynamical arrest occurs. The dynamical trapping of the tagged particle within a cage formed by its nearest neighbors is characterized by rising barrier in $F_{dyn}(r)$. A particle cage radius is $r_{cage}\approx1.3-1.5d$ determined by the radial distribution function, $g(r)$. Key physical quantities of the local dynamics are a localization length, $r_L$, a barrier position, $r_B$, a jump distance, $\Delta r= r_B-r_L$, and a local barrier, $F_B$. Then, the harmonic curvatures $K_0$ and $K_B$ at $r_L$ and $r_B$, respectively, can be calculated.

Collective dynamic effects of particles beyond the cage on an escaping process of the tagged particle are distinct but strongly correlated to the local dynamics. Particles in the first shell need to reorganize to create extra space for hopping. This molecular reorganization triggers a displacement field, $u(r)$, which is nucleated from the cage surface and radially propagates through the remaining medium. Thus, collective particles are vibrated as harmonic oscillators with a spring constant $K_0$. By using continuum mechanics analysis, the displacement field is given by  $u(r)=\Delta r_{eff}r_{cage}^2/r^2$ for $r\geq r_{cage}$, where $\Delta r_{eff}$ is an amplitude and its mathematical form is reported in Ref. \cite{11,12}. From this, the elastic energy of an arbitrary oscillator is $K_0u^2(r)/2$ and the total  elastic energy of collective particles or the elastic barrier is 
\begin{eqnarray}
F_e &=& \int_{r_{cage}}^{\infty}\rho g(r)\frac{K_0u^2(r)}{2}4\pi r^2 dr\nonumber\\
&\approx& 12\Phi K_0\Delta r_{eff}^2(r_{cage}/d)^3,
\label{eq:4-1}
\end{eqnarray}
where $\rho=6\Phi/\pi d^3$ is the number of particles per volume and $g(r)\approx 1$ for $r \geq r_{cage}$.

After obtaining the local and elastic barrier, one can calculate the structural relaxation time by Kramer's theory \cite{11,12,13}
\begin{eqnarray}
\frac{\tau_\alpha}{\tau_s} = 1+ \frac{2\pi}{\sqrt{K_0K_B}}\frac{k_BT}{d^2}\exp\left(\frac{F_B+F_e}{k_BT} \right),
\label{eq:4}
\end{eqnarray}
where $\tau_s$ is a short relaxation time scale reported elsewhere. Note that Eq. (\ref{eq:4}) provides  $\tau_\alpha(\Phi)$. Meanwhile, direct comparison with experiments requires the temperature dependence of $\tau_\alpha$. Based on previous works, we use a density-to-temperature conversion or a thermal mapping: $T=T_g+(\Phi_g-\Phi)/\beta_g\Phi_0$ with $T_g$ being the experimental glass transition temperature (defined by $\tau_\alpha(T=T_g)=100$ s in Fig. \ref{fig:3}a), $\Phi_g=0.6157$, $\beta_g=12\times10^{-4}$ $K^{-1}$, and $\Phi_0=0.5$. For indomethacin with low concentrations of graphene plasmons, $T_g \approx 315$ $K$ \cite{11}. As shown in the mainframe of Fig. \ref{fig:3}a, theoretical ECNLE predictions for $\tau_\alpha(T)$ quantitatively agree with experiment \cite{14} for a wide range of time. The relaxation is significantly sped up with heating and glassy dynamics in graphene-drug mixtures remains unchanged from that in the pure drug. 

As shown in Fig. \ref{fig:2}, under laser illumination, photothermal effects of graphene plasmon increase temperature of binary mixtures and reduce the structural relaxation at room temperature. Figure \ref{fig:3}b exhibits numerical results for $\tau_\alpha(T+\Delta T(t))$ near air-drug interface as a function of exposure time at several values of $N$. At $1200$s of light exposure, the molecular relaxation is accelerated by approximately 3 and 9 orders of magnitude for $N=5\times10^{13}$ and $2\times10^{14}$ \ce{disks/m^3}, respectively. The structural relaxation in the drug interior can be easily calculated but interfacial dynamics is strongly related to molecular surface diffusion and drug solubility.

\begin{figure}[htp]
\includegraphics[width=8cm]{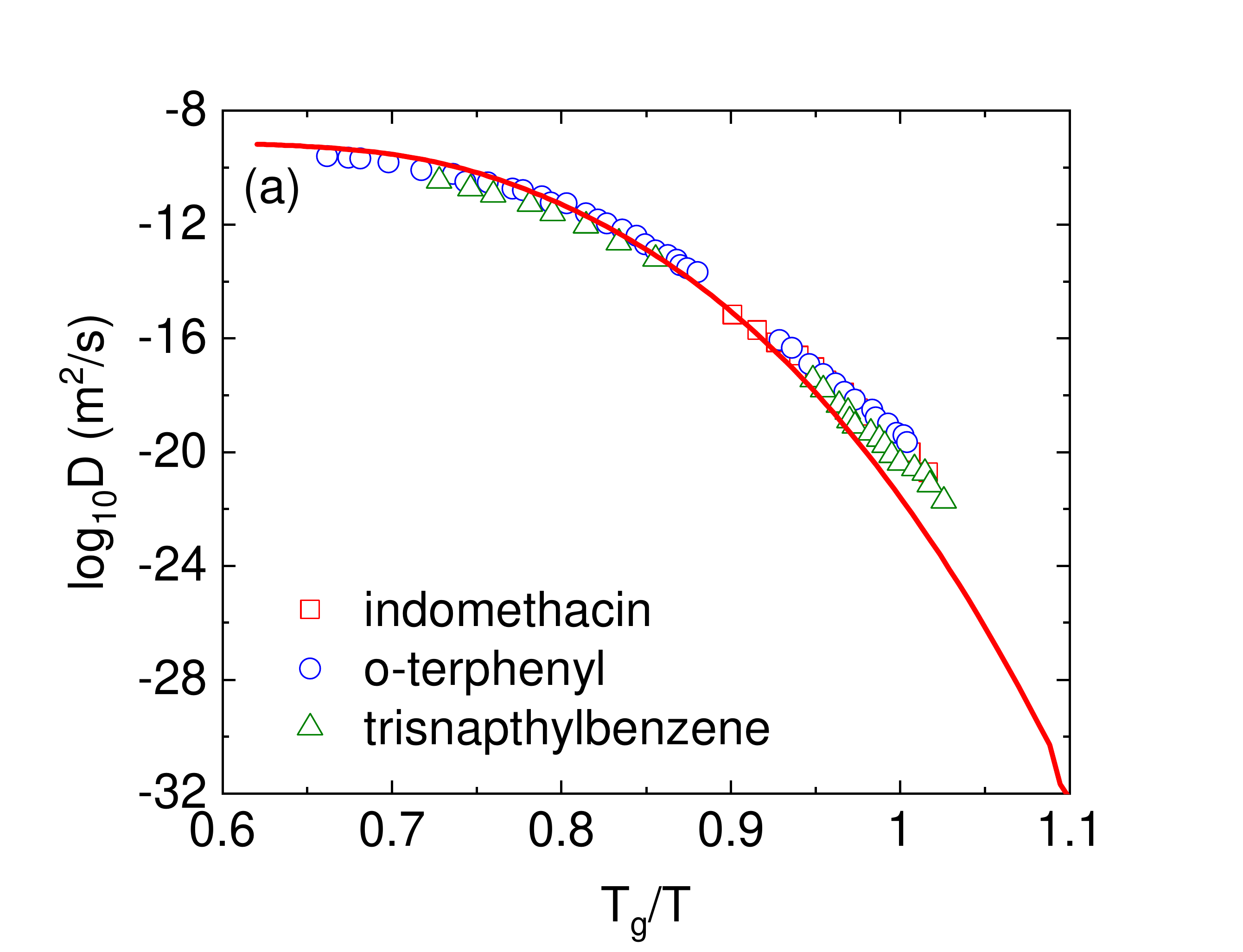}
\includegraphics[width=8cm]{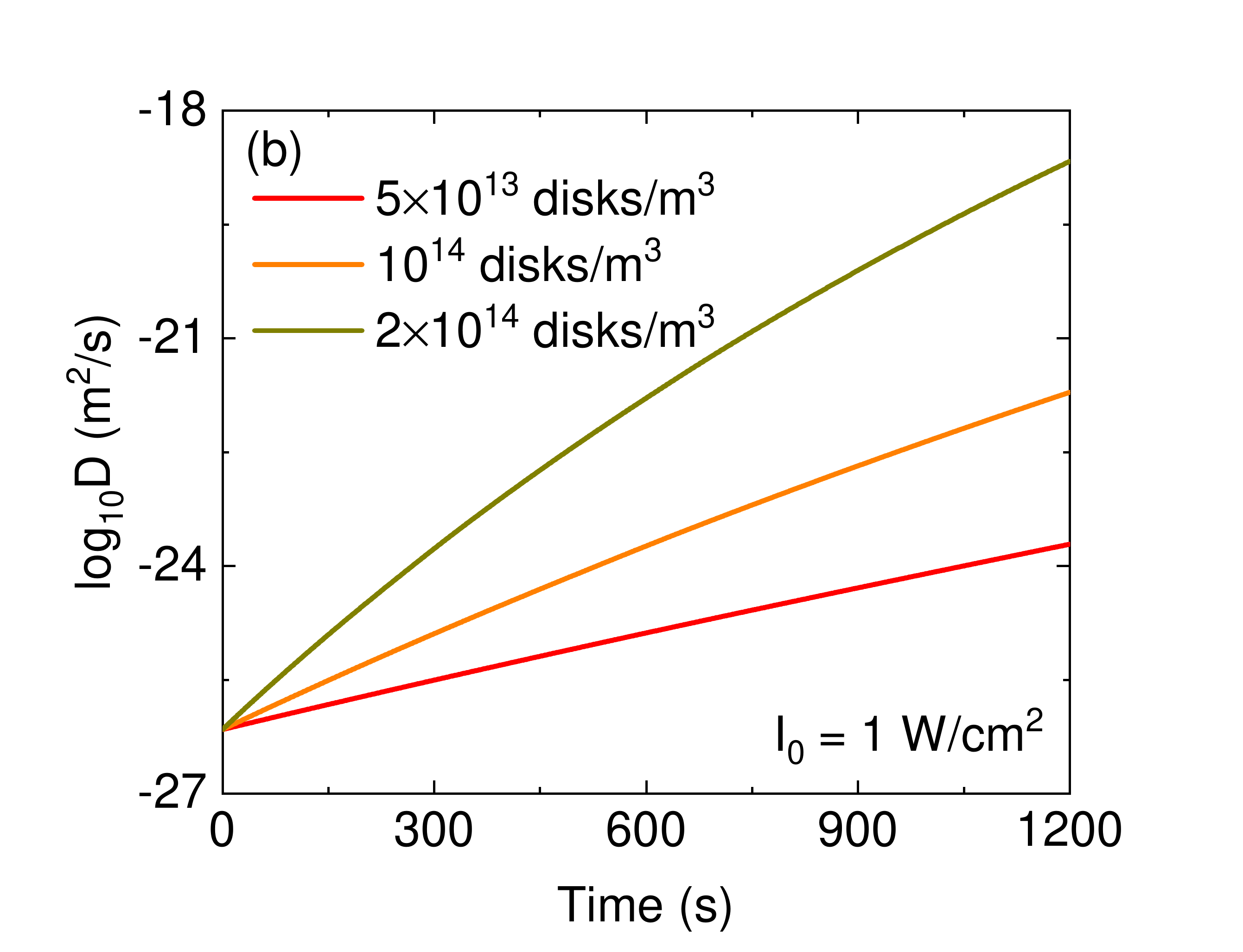}
\caption{\label{fig:4}(Color online) (a) Diffusion constant of graphene-indomethacin composite, pure indomethacin, o-terphenyl, and trisnapthylbenzene as a function of (a) temperature without laser illumination and (b) time at $T = 300K$ under laser illumination with different densities of graphene nanodisks. Data points and solid curves correspond to experimental data \cite{16} and theoretical calculations, respectively.}
\end{figure}

A relation between molecular diffusion constant and relaxation time is given by \cite{19}
\begin{eqnarray}
D(T)= \frac{\Delta r^2}{6\tau_\alpha(T)}.
\label{eq:5}
\end{eqnarray}
By using data in Fig. \ref{fig:3}a and Eq. (\ref{eq:5}), $D(T)$ for the graphene-indomethacin mixtures is predicted. We contrast theoretical prediction and experimental data of several materials including indomethacin, o-terphenyl, and trisnapthylbenzene in Fig. \ref{fig:4}a. Theory and experiment quantitatively agree with each other.

Laser-induced heating not only accelerates molecular dynamics, but also facilitates the diffusion. Numerical calculations for $D(T+\Delta T(t))$ with three densities of graphene plasmons are given in Fig. \ref{fig:4}b. Since $D(T)$ is inversely proportional to $\tau_\alpha(T)$, these physical quantities vary in the same order of magnitude. One can estimate the variation of $D(T)$ via $\tau_\alpha(T)$. In the future, the predictive diffusion constant can be used in Noyes-Whitney equation and its variants to calculate the solubility of materials.

In conclusion, we have investigated the relaxation time and diffusion of graphene-indomethacin mixtures under infrared laser irradiation. Dilute concentrations of graphene plasmons are randomly mixed in the drug. Based on the absorption cross section of individual graphene plasmons, the absorbed energy density and temperature rise are calculated when shining light. Increasing temperature makes pharmaceutical molecules move faster. Thus, the structural relaxation is sped up and the diffusion is enhanced when the photothermal heating occurs. We combine the ECNLE theory and laser-induced temperature rise prediction to quantitatively determine how plasmonic properties and density of graphene plasmons change the molecular mobility of drugs. This approach is a predictive model and could be applied to other plasmonic and amorphous materials.

\begin{acknowledgments}
This research is funded by Vietnam National Foundation for Science and Technology Development (NAFOSTED) under grant number 103.01-2019.318. Assoc. Prof. Chu Viet Ha acknowledge Projects B2019-TNA-07 of Vietnam Ministry of Education and Training.

{\bf Conflict of Interest}: The authors declare that they have no conflict of interest.

\end{acknowledgments}

\end{document}